\documentstyle[prb,multicol,aps]{revtex}
\begin{document}
\draft

\title{Electronic correlations on a metallic nanosphere. }

\author{D.N. Aristov}

\address{ Petersburg Nuclear Physics Institute,
Gatchina, St. Petersburg 188350, Russia}
\date{\today}
\maketitle

\begin{abstract}
We consider the correlation functions in a gas of electrons moving
within a thin layer on the surface of nanosize sphere. A closed form of
expressions for the RKKY indirect exchange, superconducting Cooper loop
and `density-density' correlation function is obtained. The systematic
comparison with planar results is made, the effects of spherical
geometry are outlined. The quantum coherence of electrons leads to the
enhancement of all correlations for the points--antipodes on the
sphere. This effect is lost when the radius of the sphere exceeds the
temperature coherence length.
\end{abstract}


\begin{multicols}{2} \narrowtext


Last years a considerable theoretical interest has been attracted to
the electronic properties of cylindrical and spherical nanosize
objects. This interest is mostly related to the physics of carbon
macromolecules, \cite{fulle} where the theoretical works are devoted to
the band structure calculations \cite{band} and to the effects of
topology for the trasport and mechanical properties. The discussion of
the topological aspects of electronic motion is primarily connected
with carbon nanotubes \cite{currents} and is based on the analysis of
the effective models, incorporating the particular geometry of the
object.

Other branches of the condensed matter physics, where the spherical
objects appear, concern the studies of the nonlinear optical
response in composite materials \cite{composites} and the photonic
crystals on the base of synthetic opals. \cite{opals}
In case of opals, the SiO$_2$ nanosize balls could be coated by metal
films as well. \cite{opals} The theoretical studies here are focussed
on the nonlinear optical properties and the metal coating is
characterized by an effective dielectric function. \cite{coated} This
approach should be revised when the coating has a width of a few
monolayers, \cite{Ekardt} a limit allowed by modern technologies.
\cite{SiO}

Thus the investigation of the electronic properties pertinent to the
spherical geometry can provide an interesting link between the carbon
macromolecules and other spherical structures.  On the theoretical side
of the problem, one meets a unique possibility to establish a bridge
between the methods of microscopic many-body theory and the intrinsic
spherical geometry of the atomic physics.  To be consistent, these
theoretical efforts should comprise the systematic comparison of the
quantities found for the planar and the spherical geometries of the
electronic motion. In a recent paper \cite{magnsp} we investigated the
electronic gas on a sphere in a uniform magnetic field. The exact
solution of the problem was found, two physical effects were predicted.
First is the jumps in the magnetic susceptibility at half-integer
numbers of flux quanta, piercing the sphere. Second is the localization
of the electronic states near the poles of the sphere at high fields.

In this paper we study the correlations in the electron gas moving on
the surface of the sphere. We obtain a close form of expressions for
the indirect RKKY exchange, superconducting Cooper loop,
density-density correlator. The effects peculiar to this geometry are
elucidated, their meaning for subsequent theoretical investigations is
discussed. It is particularly shown that the coherence of the
electronic motion on the sphere results in the enhancement of the
correlations between the points -- antipodes on the sphere. This effect
is lost with increasing the temperature $T$, when the temperature
coherence length becomes less than the radius of the sphere.


We consider the electrons moving on a surface of the sphere of
radius $r_0$.  The Hamiltonian of the system is given by
$     {\cal H} = -\frac{\nabla^2}{2m_e} + U(r), $
where $m_e$ is the (effective) mass of an electron and we have set
$\hbar=c=1$. The total number of electrons $N$ (with one projection of
spin) is fixed and defines the value of the chemical potential $\mu$
and the areal density $\nu = N/(4\pi r_0^2)$.  We assume that the
potential $U(r)$ confines the electrons within the thin spherical layer
$\delta r \ll r_0$, i.e. $U(r) = 0$ at $r_0< r <r_0 + \delta r$ and
$U(r) \to \infty$ otherwise.  The radial component $R(r)$ of the wave
function is a solution of the Shr\"odinger equation with the quantum
well potential.  We adopt that the chemical potential $\mu$ lies below
the first excited level of $R(r)$, which means $\delta r \lesssim \nu
^{-1/2}$.  It possible then to ignore the radial component and put
$r=r_0$ in the remaining angular part of the Hamiltonian ${\cal
H}_{\Omega}$. The eigenfunctions to the Shr\"odinger equation ${\cal
H}_{\Omega} \Psi = E\Psi$ are the spherical harmonics $Y_{lm}$ and the
spectrum is that of a free rotator model :

        \begin{equation}
        \Psi(\theta,\phi) = r_0^{-1} Y_{lm}(\theta,\phi), \quad
        E_{l} = (2m_er_0^2)^{-1} l(l+1).
        \label{spectrum}
        \end{equation}
According to (\ref{spectrum}) $\Psi(\theta,\phi)$ is normalized as
$r_0^2 \int |\Psi|^2\sin\theta d\theta d\phi  =1 $, it facilitates the
comparison of our results with the case of planar geometry.

 For the two points ${\bf r}\leftrightarrow (\theta,0)$
and ${\bf r}'\leftrightarrow (\theta',\phi)$ we define the
distance $\Omega$ on the sphere as
     $
     \cos\Omega =
     \cos\theta \cos\theta' +
     \sin\theta \sin\theta' \cos\phi
     $.
One can find \cite{magnsp} an exact representation of the electron
Green's function through the Legendre function \cite{Ba-Er}

     \begin{equation}
     G(\Omega,i\omega_n) =
     -\frac{m_e}{2\cos \pi a} P_{-1/2+a}(-\cos\Omega),
     \label{G0}
     \end{equation}
where $a = \sqrt{2m_er_0^2(\mu+i\omega_n)+1/4}$ and Matsubara frequency
$\omega_n = \pi T(2n+1)$.

In the limit $a\gg1$ and for $a\sin\Omega \gtrsim 1$ one has
\cite{magnsp} in the main order of $a^{-1}$ :

     \begin{equation}
     G(\Omega,\omega) \simeq
     - \frac{m_e}{\sqrt{2\pi a\sin\Omega}}
     \frac{\cos(a\pi-a\Omega-\pi/4)}{\cos \pi a}  ,
     \label{G0-asymp}
     \end{equation}
At the same time, if $a\gg1$ and $a(\pi-\Omega)\sim 1$ we get
     \begin{equation}
     G(\Omega\simeq\pi,\omega) \simeq
     -\frac{m_e}{2\cos \pi a} J_0\left( a(\pi-\Omega)\right),
     \label{G0-pi}
     \end{equation}
with the Bessel function $J_0(x)$. When $\Omega\to 0$, the Green's function
(\ref{G0}) diverges logarithmically, as it should.

It is convenient to introduce here the concept of the angular Fermi
momentum $L = \sqrt{2m_er_0^2 \mu+1/4} \leftrightarrow k_Fr_0 $. For
simplicity of subsequent calculations we consider the case of integer
$L$, so that at low temperatures the level $l=L-1$ is completely filled
and $l=L$ is empty. It is worth noting that the number of electrons $N=
L^2$ and the energy level spacing at the Fermi level $\Delta E \simeq
4\pi\nu/(m_eL) \simeq 2\mu/L$. For densities $\nu\sim
10^{14}\,$cm$^{-2}$ and $r_0=100\,$\AA we have $L\sim30$ and $\Delta E
\sim 10\,$meV.

The oscillating factors $\exp\pm i(a\pi- a\Omega- \pi/4)$ in
(\ref{G0-asymp}) correspond to coherence of two waves. One propagates
along the shortest way between two points and another wave goes along
the longest way, turning around the sphere.
One can show that this coherence is destroyed by the finite
quasiparticle lifetime \cite{magnsp} or by the temperature.
The latter reason takes place at $T\gtrsim \Delta E$, in
which case the Green's function acquires the form
($\omega_n>0$) :

     \begin{equation}
     G(\Omega,\pm i\omega_n) \simeq
     - \frac{m_e }{ \sqrt{2\pi L\Omega}}
     e^{ \pm i(L\Omega+\frac\pi4)
     - \omega_n \Omega/\Delta E}
     \label{G0-damped}
     \end{equation}

\noindent
This equation, after natural substitution $L\Omega\to k_Fr$,
corresponds to the usual expression for the planar geometry.



Let us now discuss the magnetic correlation function on the sphere.
This quantity defines the indirect RKKY exchange interaction between
the localized moments and thus is related to the transverse NMR
relaxation rate. It is also reponsible for the magnetic instability in
the electronic system (spin-density wave state).  The range function of
the RKKY exchange interaction is given by :

	\begin{equation}
        \chi(\Omega) = - T \sum_n G^2(\Omega,i \omega_n) .
        \label{rkk-def}
 	\end{equation}
We discuss first
the case $T\to 0 $ and use the limiting relation $T \sum_n
\to \int_{-\infty}^\infty d\omega/(2\pi)$.
Using (\ref{G0}) we write

     \begin{equation}
     \chi(\Omega) =
     -\frac{m_er_0^{-2}}{8\pi i}
     \int_C      da
     \frac{ a} {\cos^2 \pi a} \left[
     P_{a-1/2}(-\cos\Omega) \right]^2
     \label{rkk-int}
     \end{equation}
where the contour $C$ in the complex plane starts at
$e^{-i\pi/4}\infty$, passes through the point $L$ on the real axis and
goes to $e^{i\pi/4}\infty$. We can shift the integration contour to the
imaginary axis of $a$, where the integral is zero due to the oddness of
the integrand in (\ref{rkk-int}).  Upon this shift we intersect the
double poles at half-integer real $a$.  Therefore the above integral
reduces to the finite sum of residues and is equal to

     \begin{equation}
     \chi(\Omega) =
     -\frac{m_e}{4\pi^2r_0^2}
     \frac d{da} \sum_{l=0}^{L-1} \left.
     \left( a+1/2 \right)
     \left[
     P_a(-\cos\Omega) \right]^2  \right|_{a=l}.
     \label{rkk-exa}
     \end{equation}

This expression was first obtained by Larsen \cite{LarsenSph} in a
different way. Being formally exact, the expression (\ref{rkk-exa}) is
not particularly useful at $L\gg1$.  In the case of large $L$ and at
$\Omega\simeq\pi$ it was suggested in \cite{LarsenSph} to approximate
the sum (\ref{rkk-exa}) by an integral, thus obtaining

     \begin{equation}
     \chi(\Omega) \sim
     -  L \left[ P_L(-\cos\Omega) \right]^2
     \label{rkk-estim}
     \end{equation}

This estimate is however unsatisfactory at $\Omega\sim1$, since at
large $\nu$ one has $P_\nu(\cos\Omega)\sim \nu^{-1/2}
\cos(\nu\Omega-\pi/4)$. Then the leading terms in the sum
(\ref{rkk-exa}) acquire the form $\sum_l \cos2l\Omega$ and the
consideration of the next-order terms is needed.

To find $\chi(\Omega)$ in the closed form at $L\gg1$, we return back to
(\ref{rkk-def}) and use the asymptotic expressions (\ref{G0-asymp}),
(\ref{G0-pi}) for the Green's function.  Next we rescale $\omega\to2\mu
\omega/L$ (it corresponds to measuring the energy in units of spacing
$\Delta E$) and neglect the terms $\propto 1/L $ in the expansion
$a=\sqrt{L^2+2iL\omega} \simeq L+ i\omega + \omega^2/2L $.  After that
the integration over $\omega$ in (\ref{rkk-int}) is easily done and we
come to the improved estimate for (\ref{rkk-exa}) in the form :

     \begin{eqnarray}
     \chi(\Omega)
     & \simeq& -\frac{mr_0^{-2}}{4\pi^2}
      L J_0^2[L(\pi- \Omega)], \quad  (\pi-\Omega)\lesssim 1/L
     \label{impr-estimPi}
     \\   & \simeq& -
     \frac{mr_0^{-2}}{4\pi^3 \sin\Omega} \left[
     1 + \frac{ \Omega-\pi}{\sin\Omega} \sin(2L\Omega)
     \right]
     , \quad \Omega \sim 1,
     \label{impr-estim}
     \end{eqnarray}

\noindent
with the Bessel function $J_0(x)$.
One can verify that the Eqs. (\ref{impr-estimPi}) and
(\ref{impr-estim}) match smoothly at $(\pi-\Omega) \sim 1/L$.
It is convenient to rewrite (\ref{impr-estim}) further as
$\chi(\Omega)=\chi^{(1)}(\Omega)+\chi^{(2)}(\Omega)$ where

     \begin{eqnarray}
     \chi^{(1)}(\Omega) &= &
     \frac{mr_0^{-2}}{4\pi^2}
     \frac{\sin(2L\Omega)  }{\sin^2\Omega}
     \Theta\left[\frac\pi2-\Omega\right]     ,
     \label{rkk-decomp}
     \\
     \chi^{(2)}(\Omega) &= &
     - \frac{mr_0^{-2}}{4\pi^3 \sin\Omega} \left[
     1 + \frac{\sin(2L\Omega)}{\sin\Omega}
     \left[ \Omega-\pi \Theta\left[\Omega -\frac\pi2\right] \right]
     \right] ,
     \nonumber
     \end{eqnarray}
and the step function $\Theta(x)=1$ at $x>0$.
For the small angles $1/L\ll \Omega \ll 1$ the $\chi^{(1)}$
term dominates and we recover the result of the planar geometry
\cite{rkky-anyD,fnote}

     \begin{equation}
     \chi(\Omega) \simeq
     \frac{m}{(2\pi r_0\Omega)^2} \sin(2L\Omega)
     \leftrightarrow
     \frac{m}{(2\pi r)^2} \sin(2k_Fr)
     \label{rkk-planar}
     \end{equation}

Discussing the analogue of the $q-$representation of
(\ref{impr-estim}), we note that the usual formula for the
inhomogeneous static susceptibility $\chi({\bf q}) = \int d{\bf r}
e^{i{\bf qr}} \chi({\bf r})$ is replaced by $\chi_{lm} = r_0^2 \int
\chi(\Omega) Y_{lm}^\ast(\Omega,\varphi) \sin \Omega \,d\Omega
d\varphi$.  In the absence of $\varphi$ in $\chi(\Omega)$, we have
\cite{fnote2} :

        \begin{equation}
        \chi_{l} =
        2\pi r_0^2\int_0^\pi d\Omega\, \sin\Omega\,
        P_l(\cos\Omega) \chi(\Omega)
        \label{Fourier}
        \end{equation}
Recalling the property $P_l(-x) = (-1)^lP_l(x)$ and noting that
$\chi^{(2)}(\pi-\Omega) = \chi^{(2)}(\Omega)$, we conclude that the
term $\chi^{(1)}(\Omega)$ contributes to all harmonics $\chi_l$,
whereas the ``coherent'' part $\chi^{(2)}(\Omega)$ contributes only to
$\chi_l$ with even $l$. Below we show that $\chi^{(2)}$ disappears at
high enough temperatures, $T\gtrsim \Delta E$, or in the presence of
scattering.

Calculating the contribution of the term $\chi^{(1)}$, we encounter the
integral $\int_0^{\pi/2} d\Omega P_l(\cos\Omega) \sin(2L\Omega)/
\sin(\Omega)$ which is determined by small $\Omega$.  Thus one can use
$P_l(\cos\Omega)\simeq J_0[(l+1/2)\Omega]$ and extend the integration
to the infinite upper limit.

Further, the first term in $\chi^{(2)}$ yields the integral $\int_0^\pi
d\Omega P_{2l}(\cos\Omega) = \Gamma^2(l+1/2)/ \Gamma^2(l+1)\simeq
(l+\frac12)^{-1}$. The second part of $\chi^{(2)}(\Omega)$ proportional
to $\Omega\sin(2L\Omega)/ \sin^2\Omega$ is also important at
$\Omega\ll1$; we can approximate the corresponding integral by
$\int_0^\infty d\Omega J_0((l+\frac12)\Omega) \sin(2L\Omega)$.

Combining all the contributions, we have

     \begin{mathletters}
     \begin{eqnarray}
     \chi_{l} &\simeq &
     \frac{m}{2\pi} \left[
     f_1\left(\frac{l}{2L} \right)
     - \frac1{\pi L} f_2\left(\frac{l+1/2}{2L} \right)
     \right] , \\
     \chi_{l} &\simeq &
     \frac{m}{2\pi}
     f_1\left(\frac{l}{2L} \right)
     \end{eqnarray}
     \label{rkk-fou}
     \end{mathletters}

\noindent
for even and odd $l$, respectively. Here the function

     \begin{eqnarray}
     f_1(x)
     &=& \pi/2,\quad x<1   \\
     &=& \sin^{-1}(1/x), \quad x>1 \nonumber
     \end{eqnarray}
corresponds to $\chi^{(1)}(\Omega)$. It defines a cusp in $\chi_l$ at
$l=2L\leftrightarrow 2k_F$ and thus resembles the usual planar
expression. \cite{rkky-anyD} The function

     \begin{eqnarray}
     f_2(x)
     &=& x^{-1}+ (1-x^2)^{-1/2},\quad x<1  \\
     &=& x^{-1}, \quad x>1  \nonumber
     \end{eqnarray}
stems from the part $\chi^{(2)}(\Omega)$ and is present in the even
harmonics $\chi_l$.
This coherent term has a prefactor $1/L$ in (\ref{rkk-fou}) and
formally vanishes in the limit $L\propto r_0 \to \infty$.
However, $f_2(x)$ is singular at $x = 0$ and at $x=1$,
which points deserve more attention. At $l\simeq 2L$
one estimates the contribution of $\chi^{(2)}\sim L^{-1/2}$, i.e.\
much less than $\chi^{(1)}\sim 1$. At $l\sim 1$ both contributions are
comparable. Moreover, using (\ref{rkk-exa}) and relevant formulas
in \cite{Ba-Er} one can find the exact identity for
the uniform static susceptibility $\chi_{l=0} = 0 $.
%
%
The vanishing value of $\chi_{l=0}$ is connected with our choice of the
chemical potential lying between the discrete energy levels $L-1$
and $L$, so that the excited electronic states are separated from the
vacuum ones by the energy $\Delta E$. It resembles the situation with
vanishing static susceptibility in superconductors. \cite{aaa}
Exploring further this analogy, it is interesting to observe that in
the clean metal and at $T\to0$ the coherent term $f_2$ provides a
maximum of $\chi_{l}$ at $l = \sqrt2 L$. This point, similarly to the
case of ferromagnetic superconductors \cite{aaa,rkky-dwave}, could be
labeled as the momentum of helikoidal magnetic ordering.


Next, we discuss the superconducting correlation function,
which reflects the tendency to superconducting pairing and defines on
the mean-field level the superconducting temperature, $T_c$.
The basic object here (found also in the localization theory) is the
static Cooper loop, which could be written in $r-$representation as

	\begin{equation}
        \Pi(\Omega) =
        T \sum_n G(\Omega,i\omega_n) G(\Omega,-i\omega_n),
        \label{loop}
 	\end{equation}

Unlike the above case of the magnetic loop $\chi(\Omega)$, we could not
arrive at some closed type of expression for $\Pi(\Omega)$ with the use
of the exact formula (\ref{G0}) for $G(\Omega,\omega)$.  However, one
can find an approximate form of $\Pi(\Omega)$ exploring the large $L$
asymptote (\ref{G0-asymp}) for the Green's function. A calculation
similar to the above one gives the result (cf.\ (\ref{impr-estim}))

        \begin{mathletters}
        \begin{eqnarray}
        \Pi(\Omega)
        & \simeq& \frac{mr_0^{-2}}{4\pi^2}
        L J_0^2[L(\pi-\Omega)], \quad (\pi-\Omega) \lesssim 1/L
        \\   & \simeq&
        \frac{mr_0^{-2}}{4\pi^3 \sin\Omega} \left[
        \frac{\pi- \Omega}{\sin\Omega} - \sin(2L\Omega)
        \right]
        , \quad \Omega \sim 1
        \end{eqnarray}
        \label{sc-estim}
        \end{mathletters}

\noindent
Again, Eqs. (\ref{sc-estim}a)  and (\ref{sc-estim}b) smoothly match at
$(\pi-\Omega) \sim 1/L$. As above, we decompose $\Pi(\Omega)$
into the sum $\Pi^{(1)}(\Omega) + \Pi^{(2)}(\Omega)$ with

        \begin{eqnarray}
	\Pi^{(1)}(\Omega)  &=&
        \frac{mr_0^{-2}}{4\pi^2 \sin^2\Omega} \,
        \Theta\left[\frac\pi2-\Omega\right] ,
        \label{sc-decomp}
	\\
	\Pi^{(2)}(\Omega)  &=&
        \frac{mr_0^{-2}}{4\pi^3 \sin\Omega}
	\left[  \frac{
	\pi\Theta\left[\Omega-\pi/2\right]
	- \Omega }{\sin\Omega} - \sin(2L\Omega)
        \right]
	\nonumber
        \end{eqnarray}

\noindent
The property $\Pi^{(2)}(\pi- \Omega)= -\Pi^{(2)}(\Omega)$ should be
noted here.  At $1/L\ll \Omega \ll 1$ the term $\Pi^{(1)}(\Omega)$ is a
principal one and the planar result \cite{SC-anyD} is restored

     \begin{equation}
     \Pi(\Omega) \simeq
     \frac{ mr_0^{-2}}{4\pi^2 \sin^2\Omega}
     \leftrightarrow
     \frac{m}{(2\pi r)^2}
     \label{SC-planar}
     \end{equation}

Finding the momentum representation of (\ref{sc-estim}b) according to
(\ref{Fourier}), we encounter the logarithmic singularity at
$\Omega\to0$ and all $l$, which originates from the term
$\Pi^{(1)}(\Omega)$. This divergence should be cutoff at the lowest
allowable angles $\Omega_0 \sim a_0/r_0 $ with the interatomic spacing
$a_0$.

At the same time, the part $\Pi^{(2)}(\Omega)$ provides a convergent
integral which contributes only to odd harmonics, in view of the
abovementioned property. Further, it saturates at small $\Omega$, where
$\Pi^{(2)}(\Omega)$ coincides with $\chi^{(2)}(\Omega)$.

The whole result is then given by :

     \begin{mathletters}
     \begin{eqnarray}
     \Pi_{l} &\simeq &
     \frac{m}{2\pi}
     \ln\left(\frac{r_0}{a_0l} \right), \\
     \Pi_{l} &\simeq &
     \frac{m}{2\pi} \left[
     \ln \left(\frac{r_0}{a_0l} \right)
     - \frac1{\pi L} f_2\left(\frac{l+1/2}{2L} \right)
     \right] ,
     \end{eqnarray}
     \label{sc-fou}
     \end{mathletters}

\noindent
for even and odd $l$, respectively. Comparing it to (\ref{rkk-fou}) we
see that the same coherent term $f_2(l/2L)$ contributes the
harmonics of different parity. A simple analysis shows that the
presence of $f_2(l/2L)$ in (\ref{rkk-fou}) does not modify an overall
character of $\Pi_l$ for all $L\gtrsim 1$, and it is the logarithmic
term that defines the tendency to the superconducting pairing at $l\to
0$.  It should be also stressed that the static Cooper loop has a
singularity at an analog of $2k_F$, the feature absent in the planar
geometry.


Let us discuss now the `density-density' correlation function of
the electronic gas on the sphere. This function describes the variation
in the electronic density, caused by impurities, and can be written as
\cite{GoKa}

        \begin{eqnarray}
        {\cal C}({\bf r},{\bf r}') &=&
        \frac1\nu
        \langle (n({\bf r}) -\nu)
        (n({\bf r}') -\nu) \rangle -
        \delta({\bf r}-{\bf r}')
	\\ 	&=&
	- \frac1\nu
	\left( T \sum_n
        G({\bf r},{\bf r}',i\omega_n) e^{-i\omega_n \tau}
        \right)^2,
	\label{dd-cor}
        \end{eqnarray}
where $\tau \to +0$ in the last line.

In the planar geometry
we have at $T=0$ :

        \begin{eqnarray}
        {\cal C}({\bf r},{\bf r}') &=&
        -\frac1{\pi r^2} J_1^2(k_Fr)
        \simeq \frac{\sin(2k_Fr)-1}{\pi^2 k_Fr^3},
        \label{dd-2D}
        \end{eqnarray}
with $r = |{\bf r} - {\bf r}'|$. We see that at large $r$ the Friedel
oscillations with a period $(2k_F)^{-1}$ take place.

For the spherical geometry and large $L$ we use (\ref{G0-asymp}),
(\ref{dd-cor}), to obtain in the limit $T=0$ :

        \begin{eqnarray}
        {\cal C}({\bf r},{\bf r}') &=&
        \frac1{4\pi^2r_0^2L}
        \frac{\sin(2L\Omega)-1}{\sin\Omega \sin^2(\Omega/2)}
        \label{dd-sphere}
        \end{eqnarray}

This expression holds at $L\sin\Omega\gtrsim1$ and has the obvious
correspondence with (\ref{dd-2D}) for $\Omega\lesssim1$. At large
$\Omega\simeq\pi$ the amplitude of the correlations grows, in
accordance with the previous formulas (\ref{impr-estim}),
(\ref{sc-estim}).


Now let us discuss the case of finite temperatures. First, we note that
for the considered case of the fixed number of particles, the position
of the chemical potential depends on $T$ and is determined from the
equation $N = \sum_l (2l+1) n_F(E_l)$, with Fermi function $n_F(x)$.
Some analysis shows, however, that the value of $\mu$
at finite $T$ does not deviate much from its zero-temperature value
$\mu_0$. \cite{fnote3} Specifically, this deviation satisfies the
inequalities $|\mu-\mu_0|/T \lesssim 1$ and $|\mu-\mu_0| < \Delta E/2$
and is ignored below. \cite{fnote4}

The above correlation functions are not essentially modified at $T\ll
\Delta E$. The changes occur when $T\gtrsim \Delta E$, in which case we
use Eq.\ (\ref{G0-damped}) and perform the Matsubara sums in
(\ref{rkk-def}), (\ref{loop}), (\ref{dd-cor}).  As a result, we arrive
at the following expressions :

        \begin{eqnarray}
        \chi(\Omega,T) &\simeq&
        \frac{m  \sin(2L\Omega)}{(2\pi r_0\Omega)^2}
        {\cal F}\left( \frac{2\pi T\Omega}{\Delta E} \right)
        \\
        \Pi(\Omega,T) &\simeq&
        \frac{ mr_0^{-2}}{4\pi^2 \Omega^2}
        {\cal F}\left( \frac{2\pi T\Omega}{\Delta E} \right)
        \label{finiteT}
        \\
        {\cal C}({\bf r},{\bf r}',T) &\simeq&
        \frac{\sin(2L\Omega)-1}{\pi^2r_0^2L \Omega^3}
        {\cal F}^2\left( \frac{\pi T\Omega}{\Delta E} \right)
        \end{eqnarray}

\noindent
with ${\cal F}(x) = x/\sinh x$. Therefore the temperatures
$T\gtrsim \Delta E$ lead to the exponential decrease of the
correlations at $\Omega\gtrsim \Delta E/(\pi T) = \xi_T/r_0$,
with the temperature coherence length $\xi_T = 2\mu/(\pi k_F T)$.
One should also note the disappearance of the coherent terms
$\chi^{(2)}(\Omega)$ and $\Pi^{(2)}(\Omega)$ present at
$\Omega \ll 1$ in magnetic and superconducting correlators,
(\ref{rkk-decomp}) and (\ref{sc-decomp}), respectively.

Summarizing, we find a closed form of the correlation functions for the
electron gas on the sphere.
The effects peculiar to this geometry are analyzed, the role of finite
temperatures for our results is elucidated.


I thank S.L. Ginzburg, S.V. Maleyev, A.G. Yashenkin, A.V. Lazuta
for stimulating discussions.
The partial financial support from the Russian State Program for 
Statistical Physics (Grant VIII-2), grant FTNS 99-1134 and grant 
INTAS 97-1342 is gratefully acknowledged.

\end{multicols}


\begin{references}

\bibitem{fulle} see e.g. M. Dresselhaus, G. Dresselhaus, and P.C.
Eklund, {\em Science of Fullerenes and Carbon Nanotubes} (Academic
Press, San Diego, 1996).

\bibitem{band} X. Blase
{\em et al.}, Phys.Rev.Lett. {\bf72}, 1878 (1994) and references
therein.

\bibitem{currents}
P.J.Lin-Chung and A.K. Rajagopal, Phys. Rev. B {\bf 49}, 8454 (1994) ;
Y. Miyamoto, S.G.Louie, M.L. Cohen, Phys.Rev.Lett. {\bf78}, 2811 (1997).

\bibitem{composites} N.A. Nicorovici, R.C. McPhedran, G.W. Milton,
Phys. Rev. B {\bf 49}, 8479, (1994) ;
K.W.Yu, P.M.Hui, D.Stroud, Phys. Rev. B {\bf 47}, 14150, (1993).

\bibitem{opals}
S.G.\ Romanov {\em et al.}, Appl.Phys.Lett. {\bf70}, 2091 (1997);
Yu.A.\ Vlasov {\em et al.}, Phys.Rev. B {\bf55}, R13357 (1997).

\bibitem{coated}
N. Kalyaniwalla {\em et al.}, Phys.Rev. A {\bf42}, 5613 (1990);
D.J. Bergman {\em et al.}, Phys.Rev. B {\bf49}, 129 (1994);
Liang Fu and L. Resca, Phys. Rev. B {\bf 56}, 10963 (1997).

\bibitem{Ekardt} W.Ekardt, Phys. Rev. B {\bf 34}, 526, (1986).

\bibitem{SiO} R.C. Salvarezza
{\em et al.}, Phys.Rev.Lett. {\bf77}, 4572 (1996).

\bibitem{magnsp} D.N. Aristov, 
Phys.Rev. B {\bf 59}, 6368 (1999).

\bibitem {Ba-Er} H. Bateman, A. Erd\'elyi,
{\em Higher Transcendental Functions}, vols.1--2,
(McGraw -- Hill, New York, 1953).

\bibitem{LarsenSph} U. Larsen, Phys. Lett. A {\bf 122}, 361 (1987).

\bibitem{rkky-anyD} D.N. Aristov, Phys. Rev. B {\bf 55}, 8064, (1997).


\bibitem{fnote} The Eq.(10) in \cite{rkky-anyD} contains a misprint, it
should read as $ \chi(R) \propto - \sin( 2k_FR - \pi n /2 )$.

\bibitem{fnote2} We omit the factor $\sqrt{(2l+1)/(4\pi)}$, written in
the standard definition of $Y_{lm}(\theta,\phi)$. It corresponds to our
choice of the inverse Fourier transform as $\chi(\Omega) = \sum_l
(l+1/2) P_l(\cos\Omega) \chi_l$.

\bibitem{aaa}  A.A.Abrikosov,{\em Fundamentals of the
theory of metals}, North-\-Holland, Amsterdam (1988).

\bibitem{rkky-dwave} D.N. Aristov, S.V.  Maleyev, A.G. Yashenkin,
 Z.Phys. B {\bf 102}, 467 (1997).

\bibitem{SC-anyD} D.N. Aristov,
cond-mat/9807381.

\bibitem{GoKa} D.I.Golosov, M.I. Kaganov,
J.Phys.: Condens.Matter {\bf 5}, 1481 (1993).

\bibitem{fnote3} If the Fermi level $L-1$ is not completely filled,
$\mu_0$ is defined as the limiting value of $\mu$ at $T\to 0$.

\bibitem{fnote4}
The relative smallness of this deviation is
connected with the fact that in the planar limit $\Delta E\to 0$ we
recover the two-dimensional, and hence constant, density of states.

\end{references}
\end{document}